\documentclass[runningheads]{llncs}
\usepackage{graphicx}
\usepackage{amsmath}
\usepackage{float}
\usepackage[caption=false]{subfig}
\usepackage{makecell}
\usepackage{comment}
\usepackage[title]{appendix}
\begin{document}
\title{A Quality Metric for Visualization of Clusters in Graphs\thanks{This work is supported by ARC DP grant.}}
%
%
\author{Amyra Meidiana\inst{1} \and
Seok-Hee Hong\inst{1} \and
Peter Eades\inst{1} \and
Daniel Keim\inst{2}}
\authorrunning{A. Meidiana et al.}
%
\institute{University of Sydney, Australia \\
\email{amei2916@uni.sydney.edu.au, \{seokhee.hong, peter.eades\}@sydney.edu.au} \and
University of Konstanz, Germany \\
\email{keim@uni-konstanz.de}}
\maketitle              
\begin{abstract}
Traditionally, graph quality metrics focus on readability, but recent studies show the need for metrics which are more specific to the discovery of patterns in graphs. Cluster analysis is a popular task within graph analysis, yet there is no metric yet explicitly quantifying how well a drawing of a graph represents its cluster structure. 

We define a clustering quality metric measuring how well a node-link drawing of a graph represents the clusters contained in the graph. Experiments with deforming graph drawings verify that our metric effectively captures variations in the visual cluster quality of graph drawings. We then use our metric to examine how well different graph drawing algorithms visualize cluster structures in various graphs; the results confirm that some algorithms which have been specifically designed to show cluster structures perform better than other algorithms.

\end{abstract}
\section{Introduction}
Clustering is an important task in graph analysis. Visualization can be a useful tool in this task, where a good drawing of a network should be able to highlight important group structures within the network and allow a user to accurately answer group-level analytical tasks. To this end, a number of graph layout algorithms specifically focused on faithfully depicting clusters within a graph have been introduced.

The quality of a drawing of a graph is often measured using \textit{aesthetic criteria} which rate the readability of the visualization, such as the number of edge crossings or symmetry. However, these measures become less significant when working with large graphs (e.g.~\cite{kobourov2014crossings}). More recent work considers quality metrics more extensible to large graphs, such as \textit{shape-based metrics} which compare the original topology of a graph to one derived from the positioning of vertices in its drawing~\cite{eades2017shape}. Newly introduced is also the concept of more specific quality metrics concerned with the discovery of specific patterns with visualizations~\cite{behrisch2018quality}. Although general quality metrics are still necessary, these more specific metrics are useful when developing visualizations geared for a more specific purpose - for example, clustered graph visualizations which can be used to support various classes of group-level tasks~\cite{saket2014group}.

Despite a longstanding recognition of cluster discovery as one important goal in graph visualization and the definition of quality metrics that regard the depiction or discovery of specific structures, there is yet to be defined a metric that explicitly quantifies how well a visualization represents the underlying clustering structure of the graph. We therefore introduce a \textit{clustering quality metric} which scores a drawing of a graph based on how well the clustering structure of the graph is displayed within it. We present the following contributions:
\begin{enumerate}
    \item We define the \textit{clustering quality metric}, a new metric to measure the visual cluster quality of node-link graph drawings. In our framework, we compare the ground truth clustering provided for the vertices a graph to the geometric clustering derived from the graph's drawing, and the similarity of both clusterings denotes the quality of the visualization of clusters within the drawing.
    \item We validate the metric through deformation experiments of graph drawings. Results of the experiment confirm that as the graphs are distorted resulting in the clusters to become visually less distinct from each other in the drawings, the scores computed using our metric decrease.
    \item We compare various graph drawing algorithms using our metric to discover which methods perform better in visualizing cluster structures. We compare drawing algorithms of different types, including layouts that have been designed specifically to emphasize clusters. Our experiments confirm that these layouts perform better than others not explicitly geared towards cluster visualization, especially for real world graphs. 
\end{enumerate}

\section{Related Work}
\label{sec:litreview}

\subsection{Graph Drawing Quality Metrics}

\textit{Aesthetics} have been described as one criterion to be achieved by graph drawing algorithms~\cite{battista1998graph}. The concept of aesthetics is concerned with the readability of graphs and include standards such as the minimization of edge crossing and bends, and minimization of drawing area used. A number of studies have verified the correlation of such aesthetic metrics with the ability of users to execute tasks on the graph (e.g. ~\cite{huang2008effects,purchase1997aesthetic,purchase1995validating}). However, these studies tend to focus on smaller graphs, and newer studies (e.g. \cite{kobourov2014crossings}) have discovered that the effects of these aesthetic criteria are not as apparent in larger graphs.

\textit{Shape-based metrics}~\cite{eades2017shape} attempt to address this limitation by computing a shape graph based on the drawing of a graph, where two vertices are connected with an edge if they are ``close" to each other, and comparing it to the topology of the original graph - a good drawing is expected to have a shape graph similar to its actual topology. For recent work on visualization quality metrics, Behrisch et al~\cite{behrisch2018quality} provides a survey covering various visualization techniques, including but not limited to node-link drawings, and notes that measuring the effectiveness of node-link drawings in supporting analytical tasks is an open research question.

\subsection{Clustering Comparison Metrics}
\label{subsec:clustcomp}

\textit{Clustering} refers to the division of a set of items into \textit{clusters}, where items in the same cluster are more similar to each other than to items in a different cluster~\cite{aldenderfercluster}. Despite the seemingly simple definition, the notions of ``similarity" and what constitutes a ``cluster" differ between contexts, leading to the birth of various clustering algorithms and thus multiple ways to cluster the same set~\cite{estivill2002why}. To compare two clusterings \(C\) and \(C'\) of the same set, a number of metrics exist:
\begin{itemize}
    \item \textit{Rand Index (RI)} measures the similarity of \(C\) and \(C'\) based on the number of pairs of elements classified into the same group in both \(C\) and \(C'\) and the number of pairs of elements classified into different groups in both \(C\) and \(C'\)~\cite{rand1971objective}. \textit{Adjusted Rand Index (ARI)}~\cite{hubert1985comparing} is a version corrected for chance.
    \item \textit{Mutual Information (MI)}, when applied to two random variables, measures how much information of one can be gathered from the other, and is also applicable to comparisons between two clusterings \(C\) and \(C'\)~\cite{cover1991elements}. \textit{Normalized Mutual Information (NMI)}~\cite{strehl2002cluster} is a normalized version, while \textit{Adjusted Mutual Information (AMI)}~\cite{vinh2010information} is a version adjusted for chance.
    \item \textit{Fowlkes-Mallows Index (FMI)} compares a clustering \(C'\) to a target clustering \(C\) using the number of true positives, false positives, and false negatives~\cite{fowlkes1983method}.
    \item \textit{Homogeneity (HOM)} and \textit{completeness (CMP)} have been described as desirable outcomes of a cluster assignment \(C'\) compared to a target clustering \(C\), where homogeneity measures to what extent each cluster in \(C'\) only contains members of the same cluster in \(C\), and completeness refers to the extent that all members of a cluster in \(C\) are assigned to the same cluster in \(C'\)~\cite{rosenberg2007v}.
\end{itemize}

\subsection{Graph Drawing Algorithms}
\label{subsec:graphdraw}

In this section, we briefly describe a number of types of algorithms used to compute graph layouts:

\begin{itemize}
    \item \textit{Force-directed} layouts model a graph as a system where repulsive forces exist between all pairs of vertices and neighboring vertices attract each other~\cite{fruchterman1991graph}.
    \item \textit{Multi-level} layouts improve the time efficiency of force-directed layouts through steps of \textit{coarsening} the graph into a smaller graph such as through clustering, applying the layout on the smaller graph, and using it as an inital layout to draw the less coarse graph until a layout for the original is computed~\cite{hachul2004drawing}.
    \item \textit{Multi-dimensional scaling (MDS)} methods are based on dimension reduction techniques that aim to display high-dimensional data in fewer dimensions while preserving the distances between the data points~\cite{torgerson1952multidimensional}.
    \item \textit{Stress-based} layouts utilize the stress function found in the MDS literature. These methods compute a layout by minimizing an adapted stress function that considers the geometric and theoretical distances between vertices~\cite{gansner2004graph}.
    \item \textit{Spectral} methods computes the layout of a graph using the eigenvectors of matrices related to the graph, such as adjacency or Laplacian matrices~\cite{koren2005drawing}.
\end{itemize}

\section{Clustering Metric for Graph Visualization}
\label{sec:clustmetric}

\begin{figure}[h]
\centering
\includegraphics[width=0.9\textwidth]{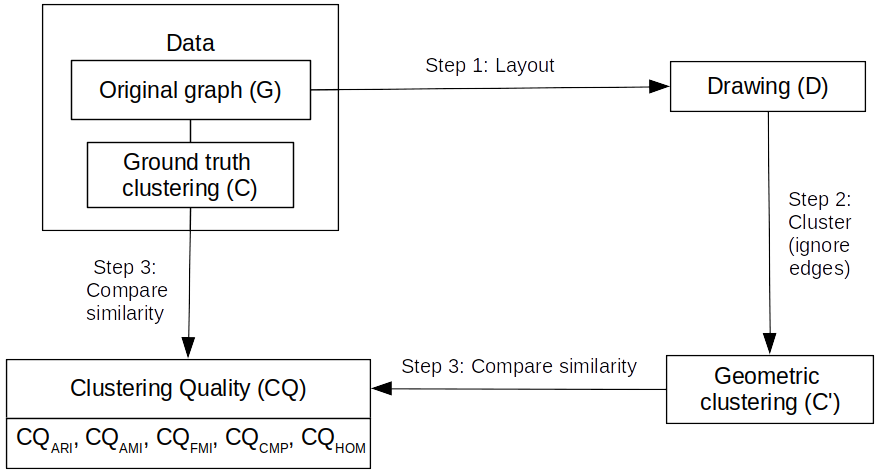}
\caption{The framework for the clustering quality metric. The framework takes as input a graph \(G\) with a predefined ground truth clustering \(C\). A drawing \(D\) is produced by applying a layout algorithm to \(G\), from which a geometric clustering \(C'\) of the vertices is computed. Computing the similarity of \(C\) and \(C'\) produces the clustering quality \(CQ\) score, which can be done using a variety of clustering comparison metrics.}
\label{fig:metricframework}
\end{figure}

We propose a new task-specific metric for graph visualization, the \textit{clustering quality metric}, for measuring how well a drawing of a graph represents its underlying clustering structure. We compute the similarity between a ground truth clustering of a graph's vertices to a geometric clustering derived from its drawing and compute the clustering quality using the similarity of the two clusterings. Figure \ref{fig:metricframework} summarizes the framework used for our proposed metric.

Let \(G = (V, E)\) be a graph and \(C = \{C_i, i=1...k\}\) be the ground truth clustering of \(V\), the vertex set of \(G\). Although in some applications a vertex may belong to multiple clusters, in this study, we focus on non-overlapping clusters as a starting point in developing the metric.

\textbf{Step 1:} We apply a layout algorithm to \(G\) to obtain a graph drawing \(D\), which provides geometric positions for each node in \(G\). A node-link drawing of a graph with no additional visual variables implicitly denotes groupings of vertices through the proximity of vertices to each other and a user is more likely to perceive two vertices drawn close together as belonging to the same group rather than two vertices drawn further apart.

\textbf{Step 2:}  We compute a geometric clustering \(C' = \{C'_i, i=1...k\}\) purely based on the geometric positions of vertices in \(D\). Any geometric clustering algorithm can be used, but in this work, we use \textit{k-means clustering}, which partitions a set into \(k\) subsets that minimize the within-class variance~\cite{macqueen1967some}. We use \(k\)-means clustering as it is a widely used method applicable to geometric clustering with existing fast and efficient heuristic approximations and because for our experiments, we know the number of ground truth clusters.

\textbf{Step 3:} Using \(C'\), we compute the clustering quality of \(D\) by computing the similarity of \(C\) with \(C'\) to produce a clustering quality score \(CQ\). Any clustering comparison metrics can be used with our framework, however we use the following metrics discussed in Section \ref{subsec:clustcomp}: Adjusted Rand Index (\(CQ_{ARI}\)), Adjusted Mutual Information (\(CQ_{AMI}\)), Fowlkes-Mallows Index (\(CQ_{FMI}\)), Homogeneity (\(CQ_{HOM}\)), and Completeness (\(CQ_{CMP}\)). These metrics have been established for measuring a clustering's quality when a target ground truth is available. In the cases of \(CQ_{ARI}\) and \(CQ_{AMI}\), they were taken over other variants of \(RI\) and \(MI\) as they are adjusted for chance. All these metrics produce a score of 1 for perfect clustering, while independent clusterings attain values close to 0.

\section{Validation Experiments}
\label{sec:validation}

\subsection{Experiment Design}

To validate our metric, we designed deformation experiments for graph drawings. We start with a drawing of a graph that displays its clusters such that the number of visible clusters and their respective sizes accurately represent the ground truth clusters and the clusters are well-separated from each other with no overlap.

We then progressively deform the drawing. In each experiment, we performed 10 steps of deformation, where in each step, the coordinates of each vertex from the previous step are perturbed by a small value in the range \([0,\delta]\), with \(\delta\) being in the range of 0.05-0.1 multiplied by the drawing area. We compute the clustering quality score and compare the scores across all steps of the deformation.

Based on the clustering comparison metrics, we expect our approach to produce scores in the range of \([0,1]\) where a higher value denotes a closer similarity between the geometrical clustering \(C'\) derived from the drawing \(D\) and the ground truth clustering \(C\). Therefore, we formulate the following hypothesis in order to validate our metric:

\subsubsection{Hypothesis 1:} The clustering quality metric scores will decrease as the graph drawings are deformed.
\\
\\
To create the initial layout, we used the Backbone layout from Visone~\cite{baur2001visone} as this layout produced drawings scoring 1 or nearly 1 on our metric for our datasets. The exception is that we used sfdp from Graphviz~\cite{ellson2001graphviz} for \(cv-many-verydense-mid\) and \(gnm-many-mid-verysparse\), where sfdp produces drawings with higher clustering quality metric scores than backbone. We used cluster comparison metrics implementations from scikit-learn~\cite{pedregosa2011scikit}.

Each dataset for our validation experiment is created by first creating a small graph. Each vertex is replaced with a larger graph of a specified internal density - each will become a cluster of the dataset. Then, each edge is replaced with inter-cluster edges with a specified external density. Table \ref{table:validationdata} shows the dataset details. \(|c|\) stands for the number of clusters and \(avg(cd)\) denotes the average internal density of the clusters, as opposed to the global density denoted in the previous column. 

Each graph is named in the format \([name]-[no. of clusters]-[internal density]-[external density]\), where we vary the parameters to increase generality. The prefixes denote the structure used to generate the clustered graph - \(c\) stands for a complete graph, \(b\) denotes a bipartite graph, \(s\) denotes a star graph, \(t\) denotes a tree, \(p\) denotes a path, \(rn\) denotes an \(r\)-regular graph, \(cv\) is a complete graph with variable cluster sizes, and \(gnm\) denotes a \(G_{n,m}\) random graph.
\begin{table}[h]
    \centering
    \caption{Validation datasets}
    \begin{tabular}{|c|c|c|c|c|c|}
        \hline
        Name & \(|V|\) & \(|E|\) & \(|c|\) & Density & \(avg(cd)\) \\ \hline
        \(c-few-verydense-mid\) & 439 & 9552 & 9 & 0.0497 & 0.399 \\ \hline
        \(s-few-verydense-dense\) & 2051 & 256108 & 10 & 0.122 & 0.400 \\ \hline
        \(t-few-dense-mid\) & 2082 & 164180 & 15 & 0.0379 & 0.400\\ \hline
        \(c-few-dense-mid\) & 898 & 31516 & 9 & 0.0391 & 0.349 \\ \hline
        \(p-few-verydense-verysparse\) & 3002 & 230055 & 15 & 0.0511 & 0.759 \\ \hline
        \(c-mid-verydense-mid\) & 815 & 18674 & 15 & 0.0563 & 0.797 \\ \hline
        \(r3-mid-dense-verysparse\) & 1773 & 53103 & 20 & 0.0338 & 0.670 \\ \hline
        \(cv-many-verydense-mid\) & 2000 & 54749 & 30 & 0.0274 & 0.788 \\ \hline
        \(r3-many-verydense-sparse\) & 3045 & 124727 & 30 & 0.0269 & 0.801 \\ \hline
        \(gnm-many-mid-verysparse\) & 2685 & 26098 & 30 & 0.00724 & 0.214 \\ \hline
    \end{tabular}
    \label{table:validationdata}
\end{table}

\subsection{Results}
\begin{figure}[h]
\centering
\subfloat[Step 0]{
\includegraphics[width=0.22\columnwidth]{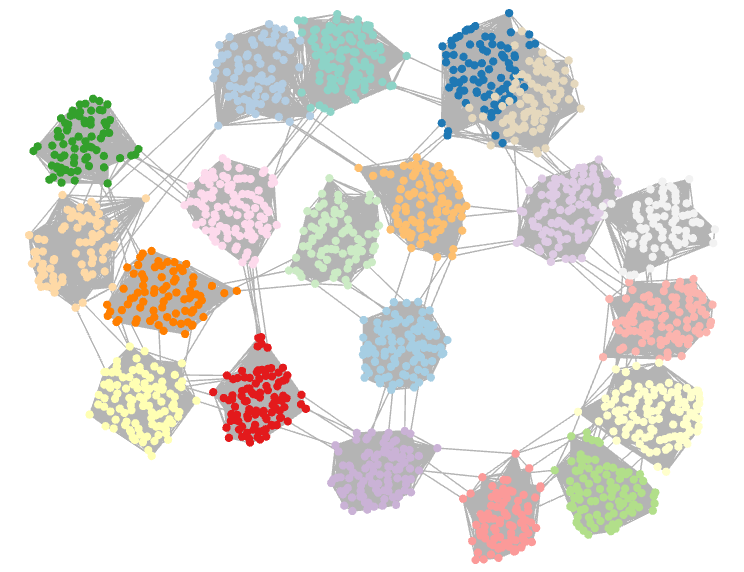}
}
\subfloat[Step 2]{
\includegraphics[width=0.22\columnwidth]{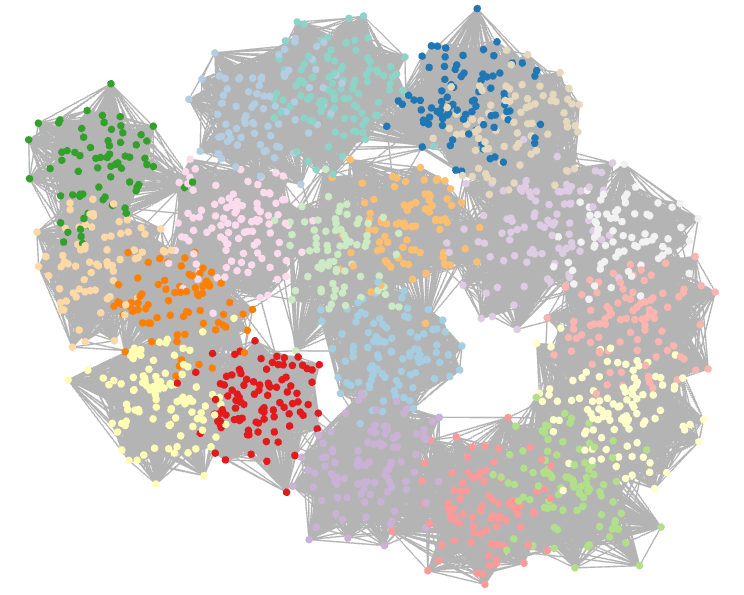}
}
\subfloat[Step 5]{
\includegraphics[width=0.22\columnwidth]{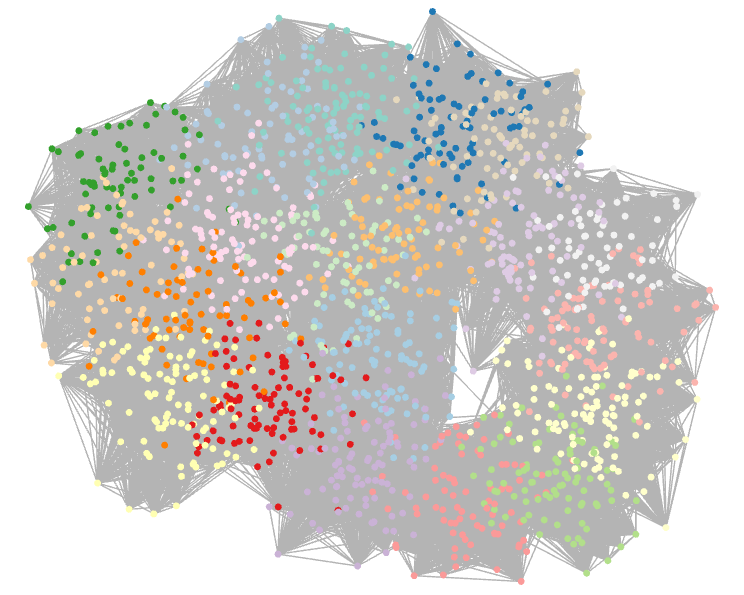}
}
\subfloat[Step 9]{
\includegraphics[width=0.22\columnwidth]{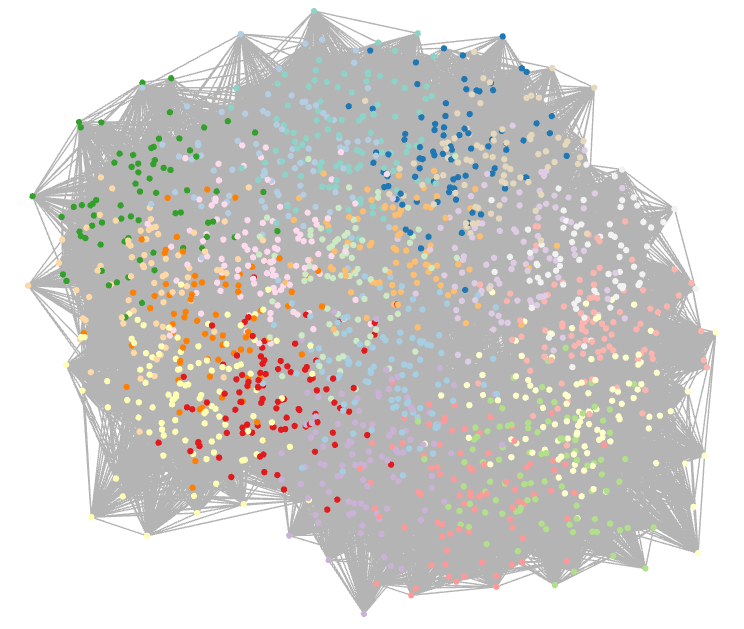}
}
\caption{Deformation experiment for \(r3-mid-dense-verysparse\), drawn using Backbone layout, showing how each subsequent step further deforms the clusters in the drawing.}
\label{fig:perturb-metrics}
\end{figure}

Figure \ref{fig:perturb-metrics} displays one deformation experiment example, where vertices are colored based on their combinatorial cluster membership. In step 0 (Figure \ref{fig:perturb-metrics} (a)), vertices of the same cluster are positioned close to each other, there is minimal overlap between each cluster and the layout produces \(CQ\) scores of 1. As the positions are perturbed, vertices of the same cluster grow further apart. The clusters also continue mixing with each other, until vertices are no less likely to be placed closer to members of other clusters than vertices in its own cluster.

\begin{figure}[h]
    \centering
    \includegraphics[width=0.8\textwidth]{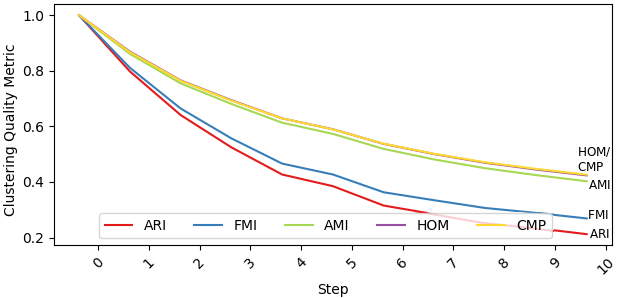}
    \caption{Average of clustering quality scores for all validation experiments. The decreasing trend for all clustering comparison metrics show that our metric successfully captures the deteriorating visual clustering quality and validates Hypothesis 1. We also see that \(CQ_{AMI}\) and \(CQ_{FMI}\) are more sensitive to changes in the visual cluster quality, from the steeper curves. Also note that \(CQ_{HOM}\) and \(CQ_{CMP}\) produce highly similar results such that their curves overlap.}
    \label{fig:metrics_perturb_average}
\end{figure}

Figure \ref{fig:metrics_perturb_average} shows the clustering metric scores for each deformation step, with the scores averaged for all datasets in Table \ref{table:validationdata}. We expect to see the \(CQ\) scores decreasing after each deformation step, which is indeed what the figure shows, confirming Hypothesis 1 for a wide variety of clustered graphs.

\subsection{Discussion and Summary}
Figure \ref{fig:metrics_perturb_average} shows that the plots of the clustering quality metric scores produce a downward slope. This validates our metric and the usage of all selected clustering comparison metrics with our framework. It can also be seen that the scores of our metric deteriorate at different rates when different clustering comparison metrics are used: \(CQ_{ARI}\) deteriorates at the fastest rate, followed closely by \(CQ_{FMI}\). \(CQ_{HOM}\) and \(CQ_{CMP}\) obtains very similar scores with their curves overlapping, while \(CQ_{AMI}\) degrades at a slightly faster rate. Therefore, we conclude that \(CQ_{ARI}\) and \(CQ_{FMI}\) are more sensitive to changes in clustering visualisation quality than the other metrics.

\textit{In summary}, the validation experiments have shown that our metric reflects the visual clustering quality of drawings of clustered graphs. Furthermore, from the different rates of change of the clustering quality scores when different clustering comparison metrics are used, we conclude that \(CQ_{ARI}\) and \(CQ_{FMI}\) are better at capturing changes in visual cluster quality and are recommended for use with our framework.

\section{Layout Comparison Experiments}
\label{sec:layoutcomp}

\subsection{Experiment Design}

After the validation experiments have shown that our metric effectively measures visual cluster quality, we compare the performance of a number of graph drawing algorithms against our metric. We selected layouts of different types:

\begin{itemize}
    \item Force-directed: \textit{Fruchterman-Reingold (FR)}~\cite{fruchterman1991graph} and \textit{Organic} from yfiles~\cite{wiese2004yfiles}.
    \item Multi-level: \textit{FM3}~\cite{hachul2004drawing} and \textit{sfdp}~\cite{ellson2001graphviz,hu2005efficient}.
    \item MDS:  \textit{Metric MDS} based on classical scaling~\cite{torgerson1952multidimensional} and \textit{Pivot MDS}~\cite{brandes2006eigensolver}.
    \item Stress-based: \textit{Stress Majorization}~\cite{gansner2004graph} and \textit{Sparse Stress Minimization}~\cite{ortmann2016sparse}.
    \item Spectral: spectral layout with graph laplacian.
\end{itemize}

We also selected a few layouts which purport to focus on the discovery of clusters or important community structures in a graph to test their claims:
\begin{itemize}
    \item \textit{LinLog}~\cite{noack2003energy} modifies the force-directed model to emphasize clusters.
    \item \textit{Backbone}~\cite{nocaj2015untangling} utilizes triadic or quadratic Simmelian backbones to extract important community structures from ``hairball" graphs.
    \item \textit{tsNET}~\cite{kruiger2017graph} is based on t-distributed Stochastic Neighbor Embedding (t-SNE), a dimensionality reduction technique~\cite{maaten2008visualizing}, and aims to preserve point neighborhoods.
\end{itemize}

Based on the selection of algorithms, we formulate the following hypothesis:

\subsubsection{Hypothesis 2:} LinLog, backbone, and tsNET will score higher on our metric than other selected layouts in visualizing clusters in graphs.
\\
\\
We used implementations provided from Tulip~\cite{david2001tulip} (FR, FM3, Pivot MDS, Stress Majorization, LinLog), visone~\cite{baur2001visone} (Backbone, Metric MDS, Sparse Stress Minimization, Spectral), yEd~\cite{wiese2004yfiles} (Organic), Graphviz~\cite{ellson2001graphviz} (sfdp), and Kruiger's implementation of tsNET~\cite{kruiger2017tsnet}. We re-used some datasets from the validation experiments and created some new ones, listed in Table \ref{table:layoutcompdata}. We also selected real world graph datasets with existing vertex categorization, which are listed under the double line in Table \ref{table:layoutcompdata}. The datasets were taken from Pajek~\cite{pajekdata} and Stanford Network Analysis Project's (SNAP) repository~\cite{snapnets,biosnapnets}. 

\begin{table}[H]
    \centering
    \small
    \caption{Additional layout comparison datasets}
    \label{table:layoutcompdata}
    \begin{tabular}{|c|c|c|c|c|c|}
        \hline
        Name & \(|V|\) & \(|E|\) & \(|c|\) & Density & avg(cd) \\ \hline
        \(b-many-dense-sparse\) & 1797 & 49210 & 30 & 0.0305 & 0.560 \\ \hline
        \(cv-mid-verydense-mid\) & 939 & 21798 & 20 & 0.0495 & 0.162 \\ \hline
        \(s-mid-mid-sparse\) & 2116 & 24175 & 20 & 0.0108 & 0.216 \\ \hline
        \(w-many-mid-verysparse\) & 2485 & 68844 & 25 & 0.0223 & 0.554 \\ \hline
        \(r4-many-verydense-verysparse\) & 3045 & 124727 & 30 & 0.0269 & 0.801 \\ \hline \hline
        \(revije-90\) & 124 & 1334 & 14 & 0.127 & 0.377 \\ \hline
        \(SS-Butterfly-0-85\) & 832 & 13009 & 10 & 0.0376 & 0.258 \\ \hline
        \(email-Eu-core-lcc\) & 986 & 16687 & 34 & 0.0344 & 0.490 \\ \hline
    \end{tabular}
\end{table}

\subsection{Results}

\begin{table}[H]
\centering
\caption{Layout comparison for \(c-few-verydense-mid\)}
\label{table:layoutcomp_cfew}
\begin{tabular}{|c|c|c|c|c|c|}
\hline
FR & Organic & Stress Maj. & Metric MDS & Backbone & FM3 \\ \hline
\includegraphics[width=0.14\textwidth]{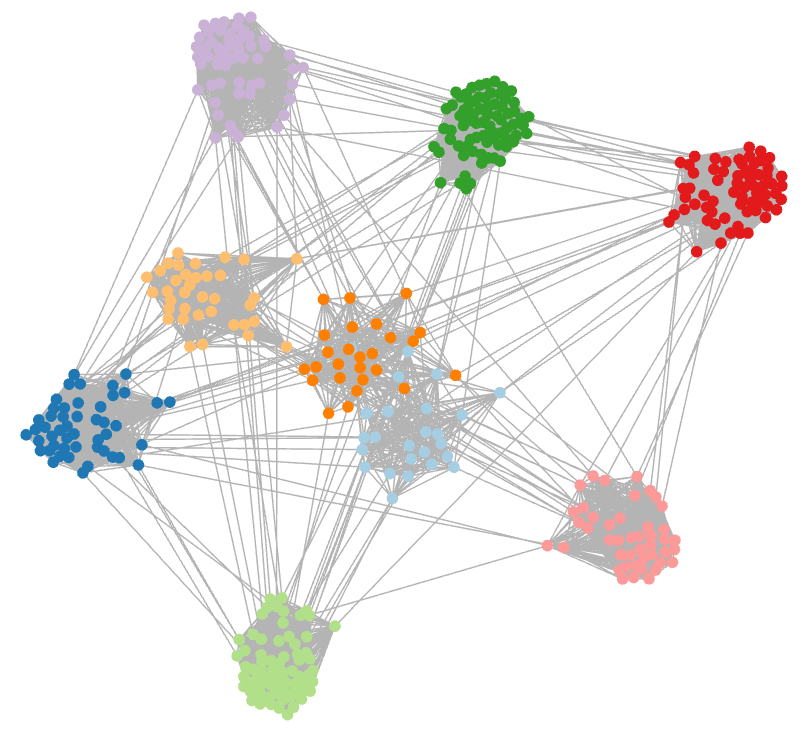} &
\includegraphics[width=0.14\textwidth]{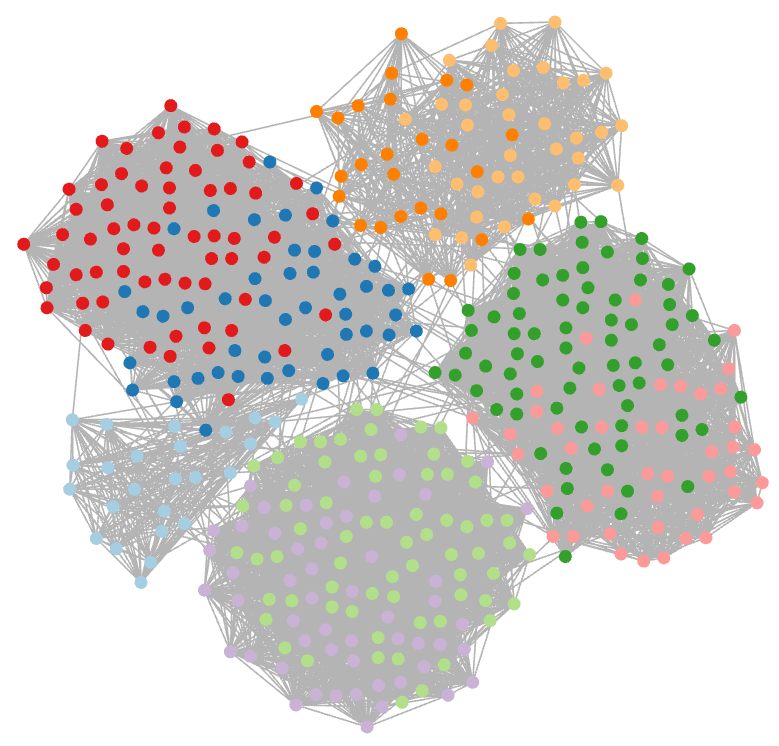} &
\includegraphics[width=0.14\textwidth]{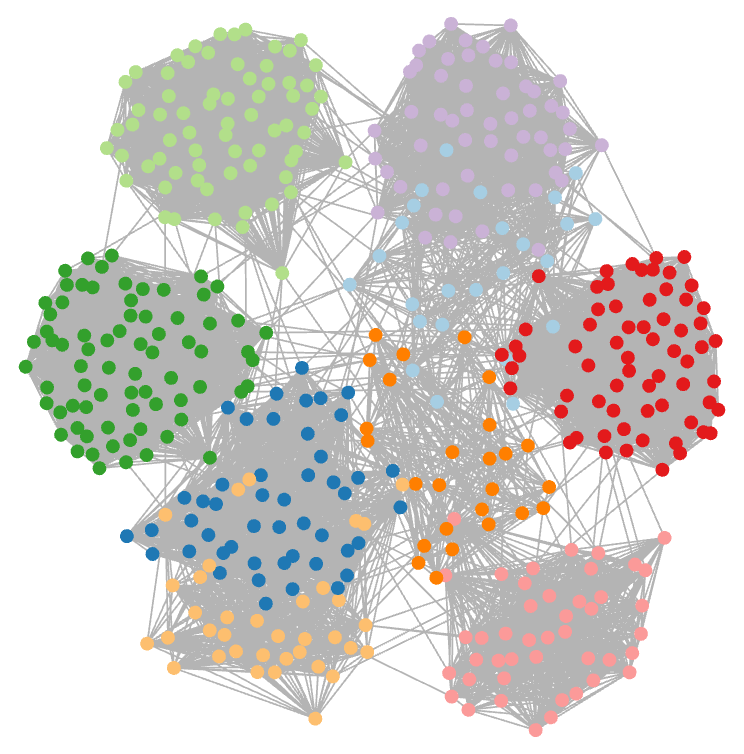} &
\includegraphics[width=0.14\textwidth]{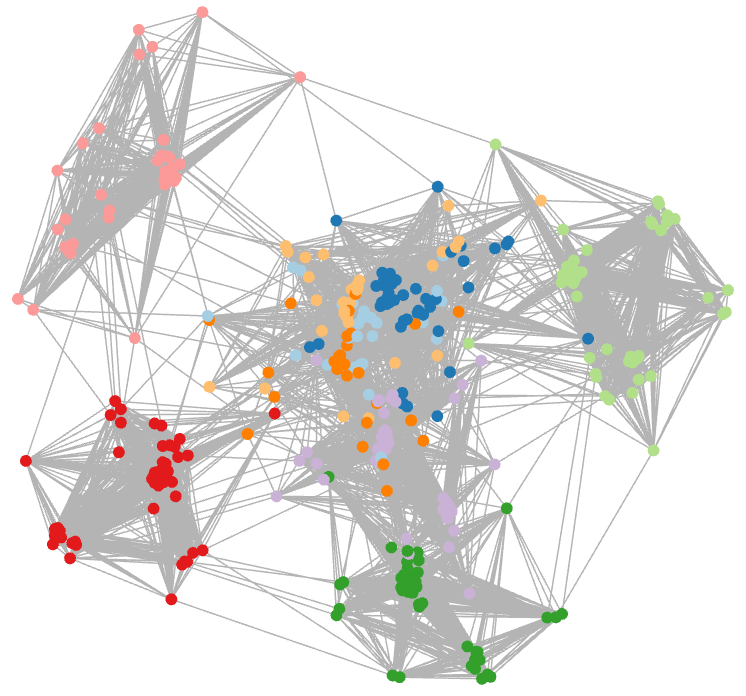} &
\includegraphics[width=0.14\textwidth]{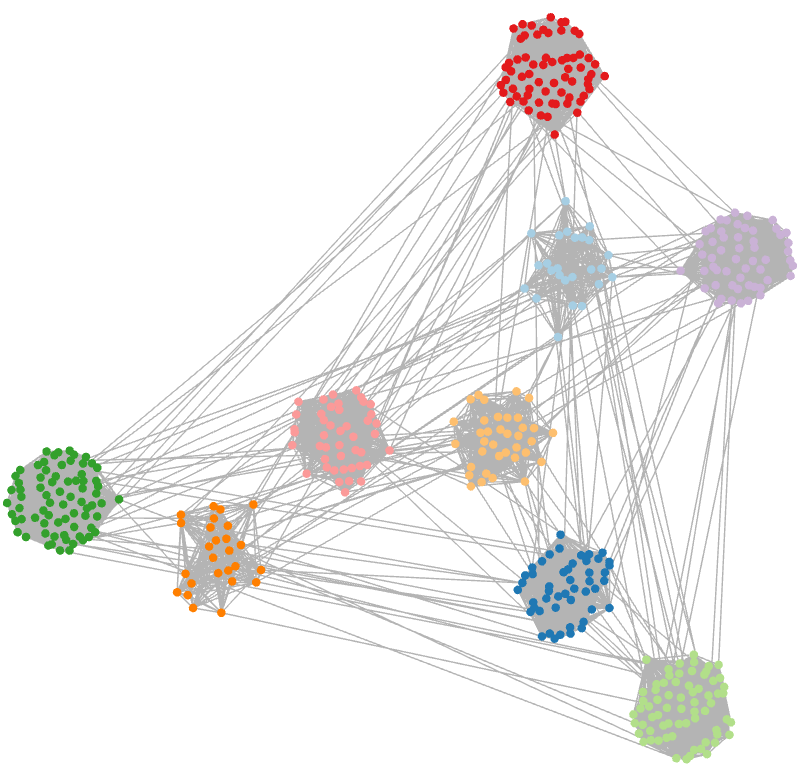} &
\includegraphics[width=0.14\textwidth]{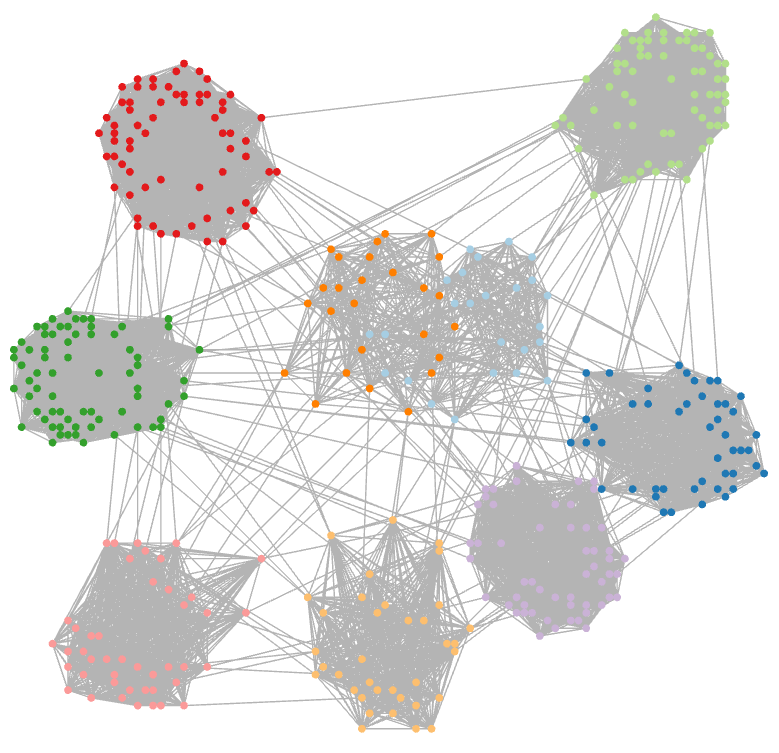}\\ \hline
Spectral & S. Stress Min. & tsNET & Pivot MDS & sfdp & LinLog \\ \hline
\includegraphics[height=0.14\textwidth]{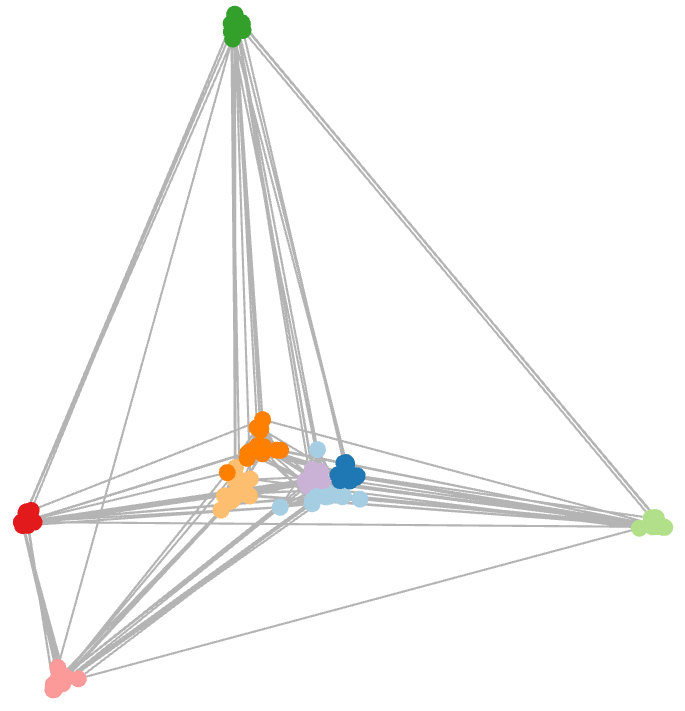} &
\includegraphics[width=0.14\textwidth]{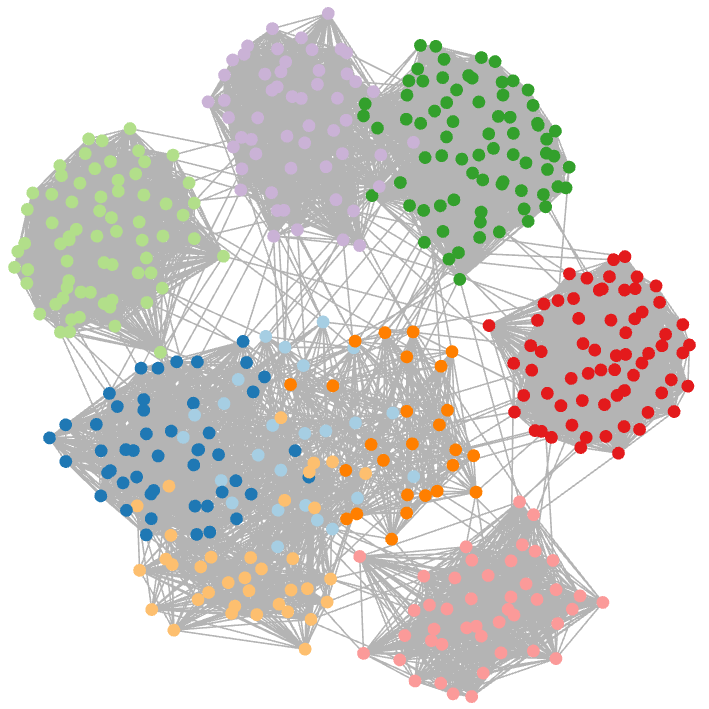} &
\includegraphics[width=0.14\textwidth]{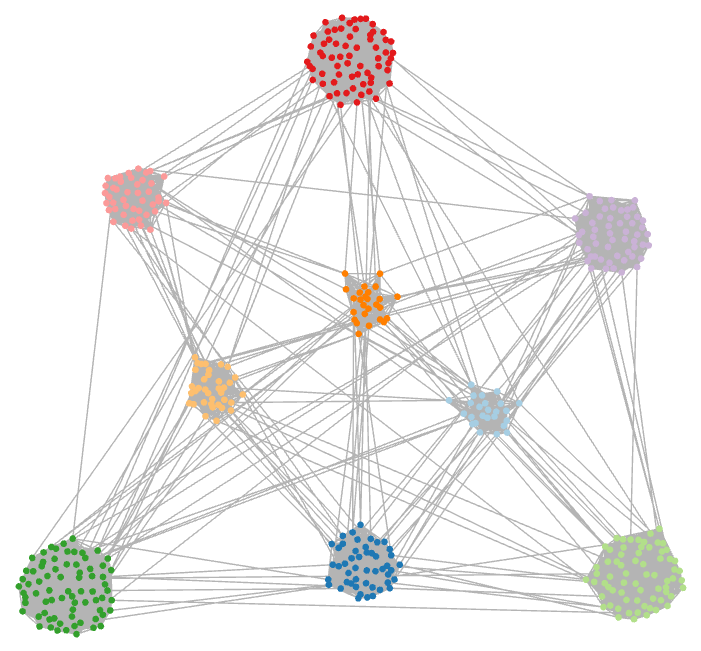} &
\includegraphics[width=0.14\textwidth]{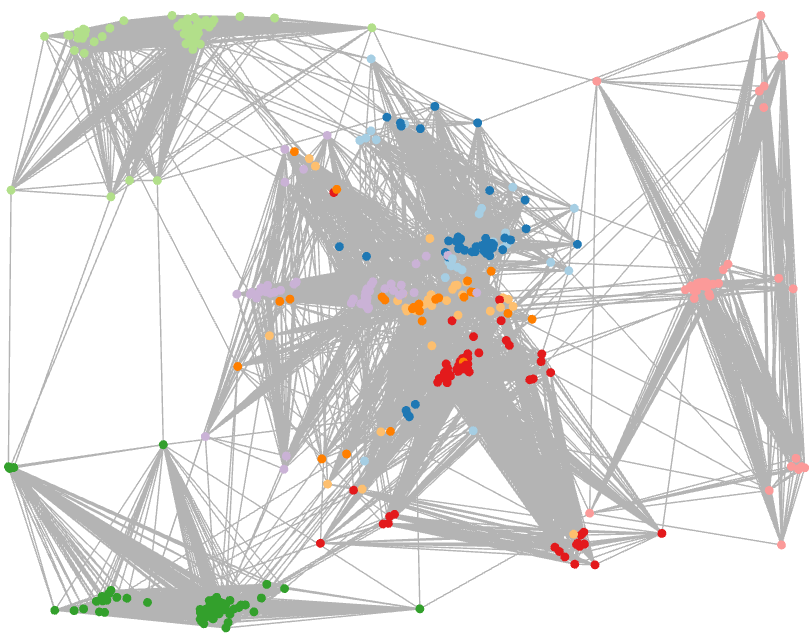} &
\includegraphics[width=0.14\textwidth]{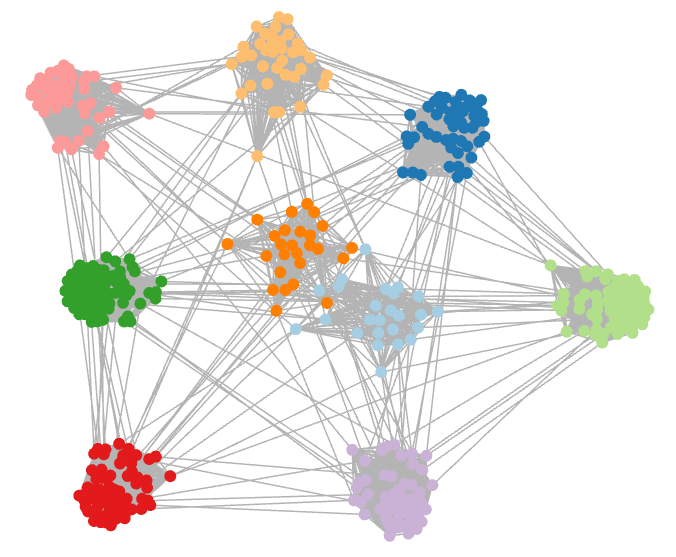} &
\includegraphics[width=0.14\textwidth]{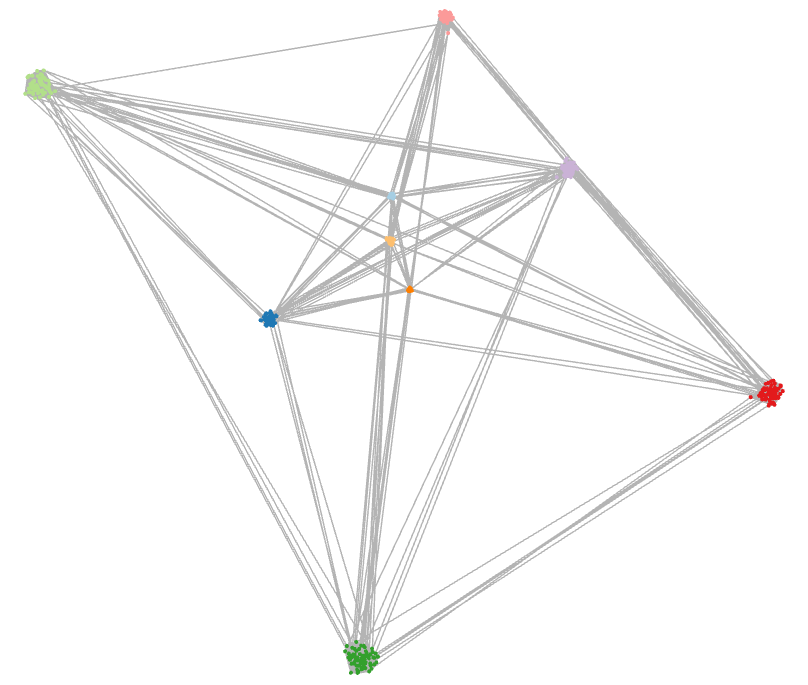} \\ \hline
\end{tabular}
\end{table}
\begin{figure}[H]
\centering
\includegraphics[width=0.8\textwidth]{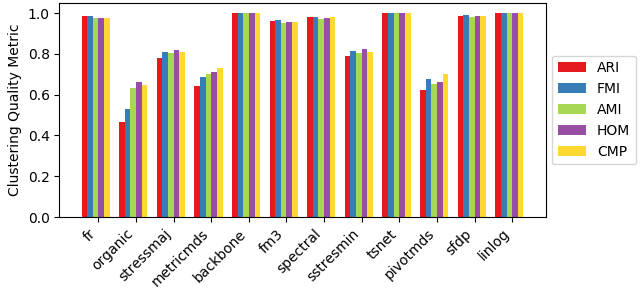}
\caption{Clustering quality metrics for \(c-few-verydense-mid\). LinLog, tsNET, and Backbone produces scores of 1 on our metrics, in line with Hypothesis 2. For this dataset, sfdp, FR, FM3, and spectral also score highly, close to 1.}
\label{fig:layoutcompmetrics_cfew}
\end{figure}

\begin{table}[H]
\centering
\caption{Layout comparison for \(email-Eu-core-lcc\)}
\label{table:layoutcomp_email}
\begin{tabular}{|c|c|c|c|}
\hline
FR & Organic & Stress Maj. & Metric MDS \\ \hline
\includegraphics[width=0.22\textwidth]{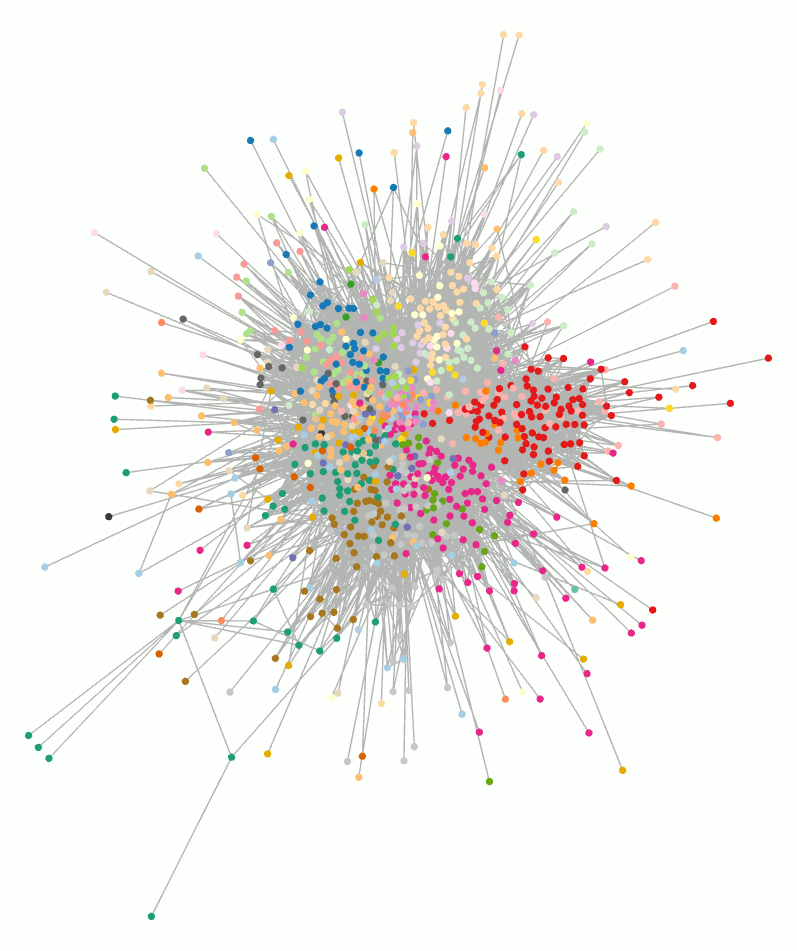} &
\includegraphics[width=0.22\textwidth]{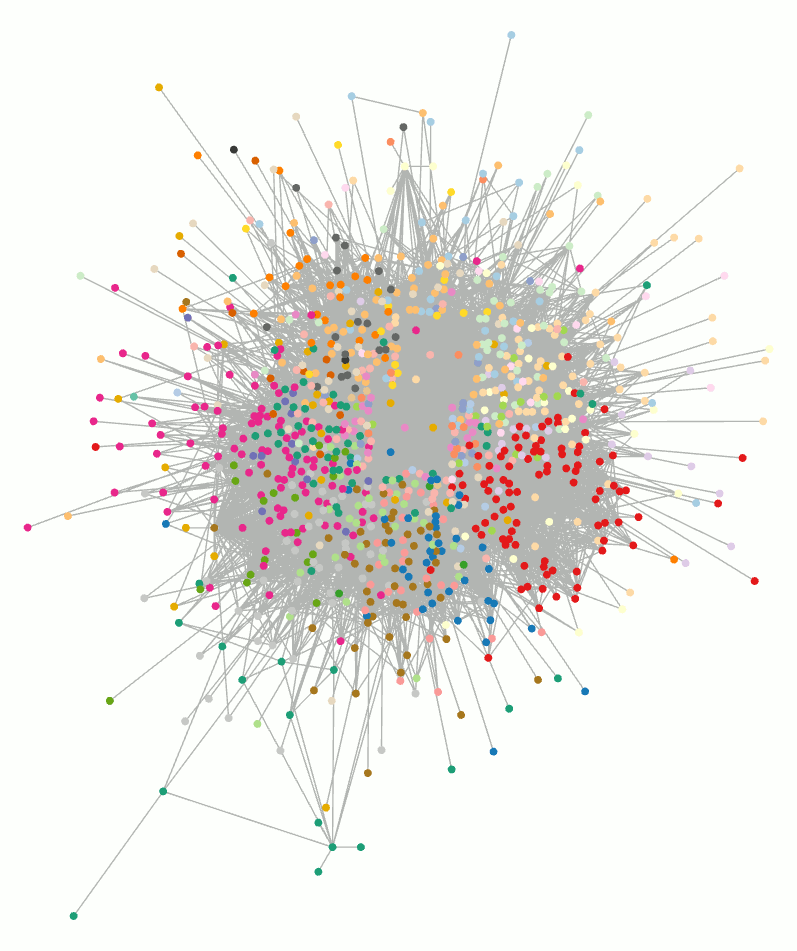} &
\includegraphics[width=0.22\textwidth]{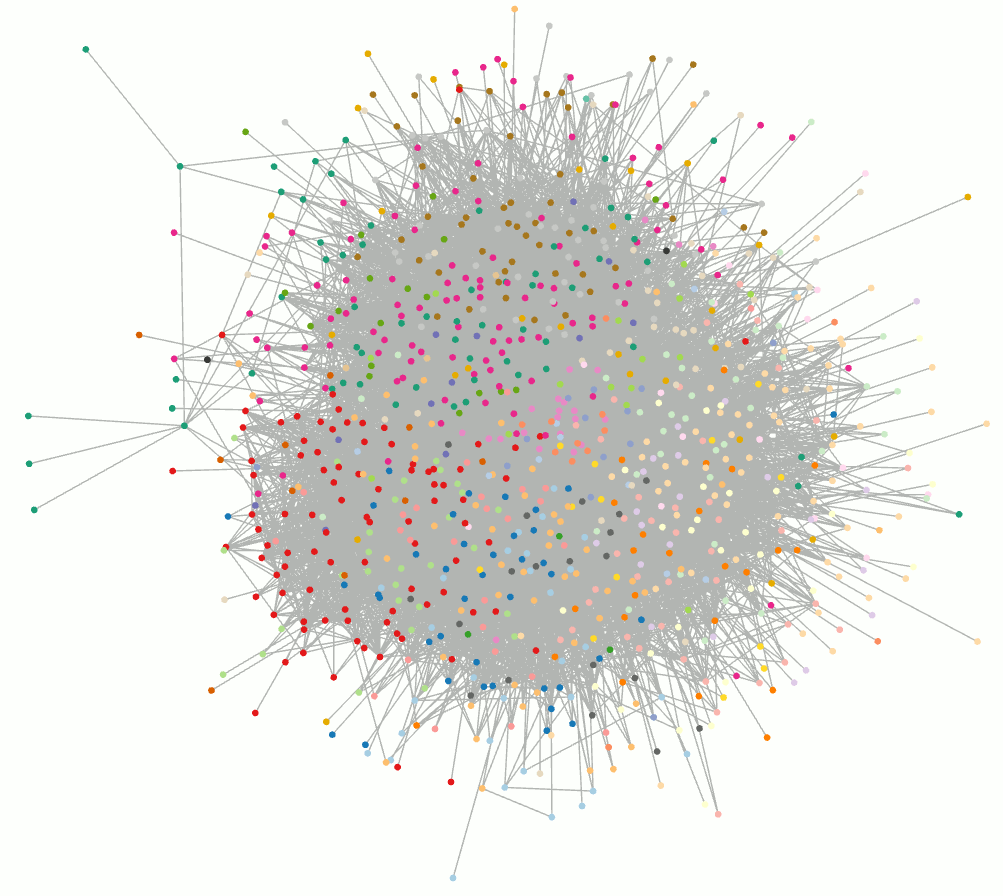} &
\includegraphics[width=0.22\textwidth]{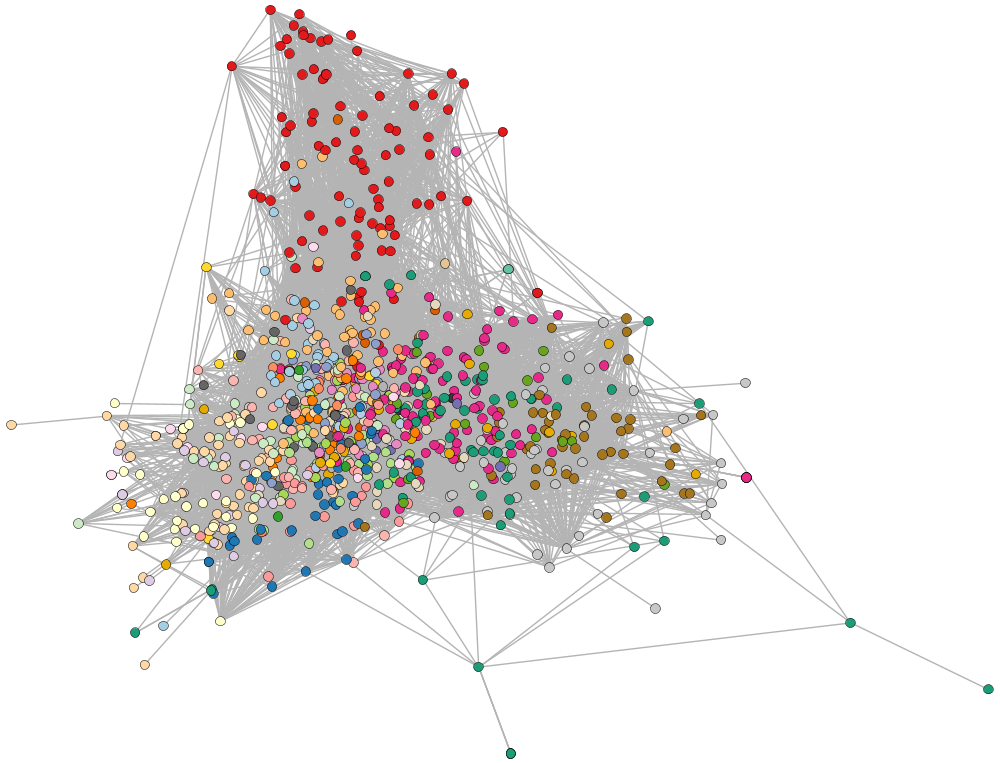} \\ \hline
Backbone & FM3 & Spectral & S. Stress Min. \\ \hline
\includegraphics[width=0.22\textwidth]{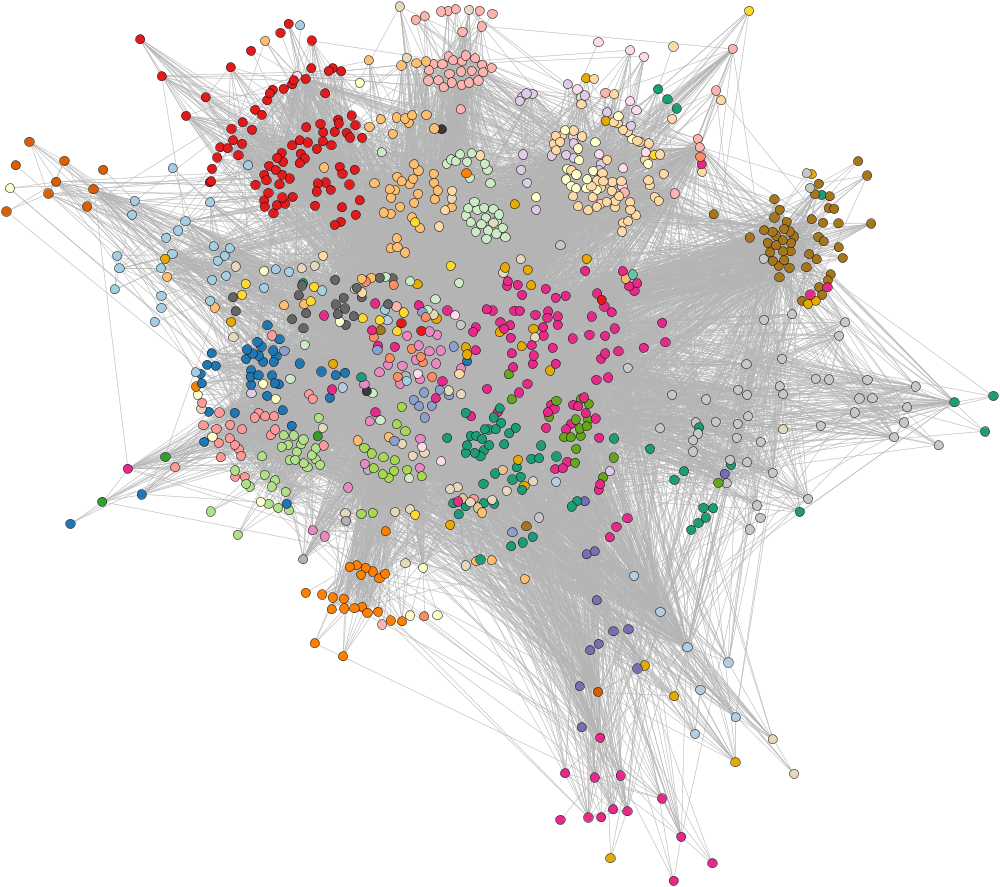} &
\includegraphics[width=0.22\textwidth]{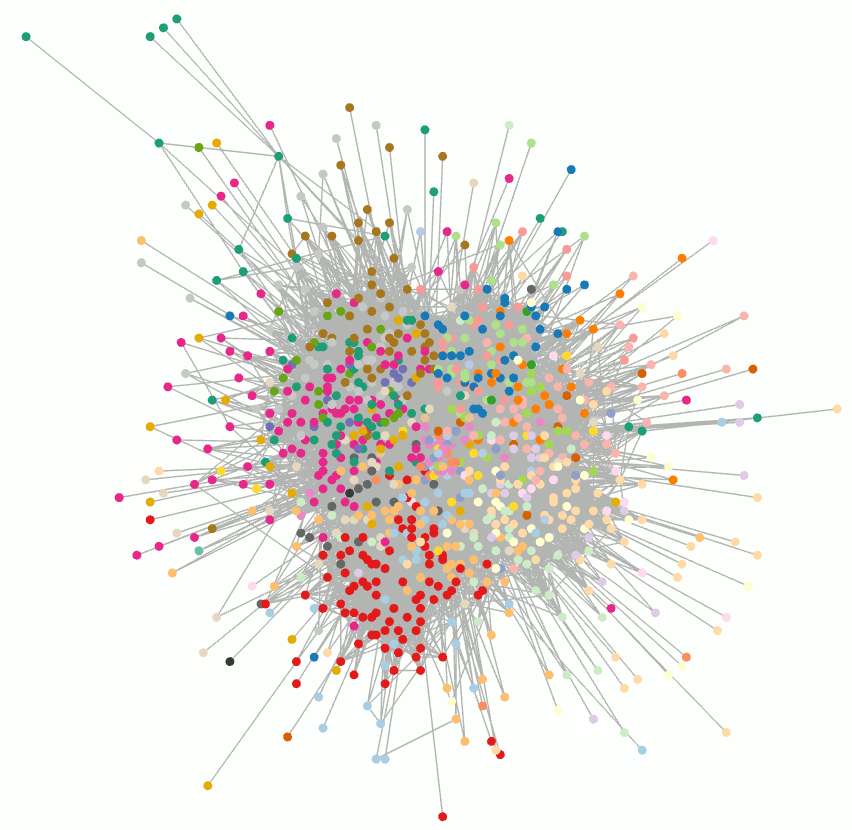} &
\includegraphics[height=0.22\textwidth]{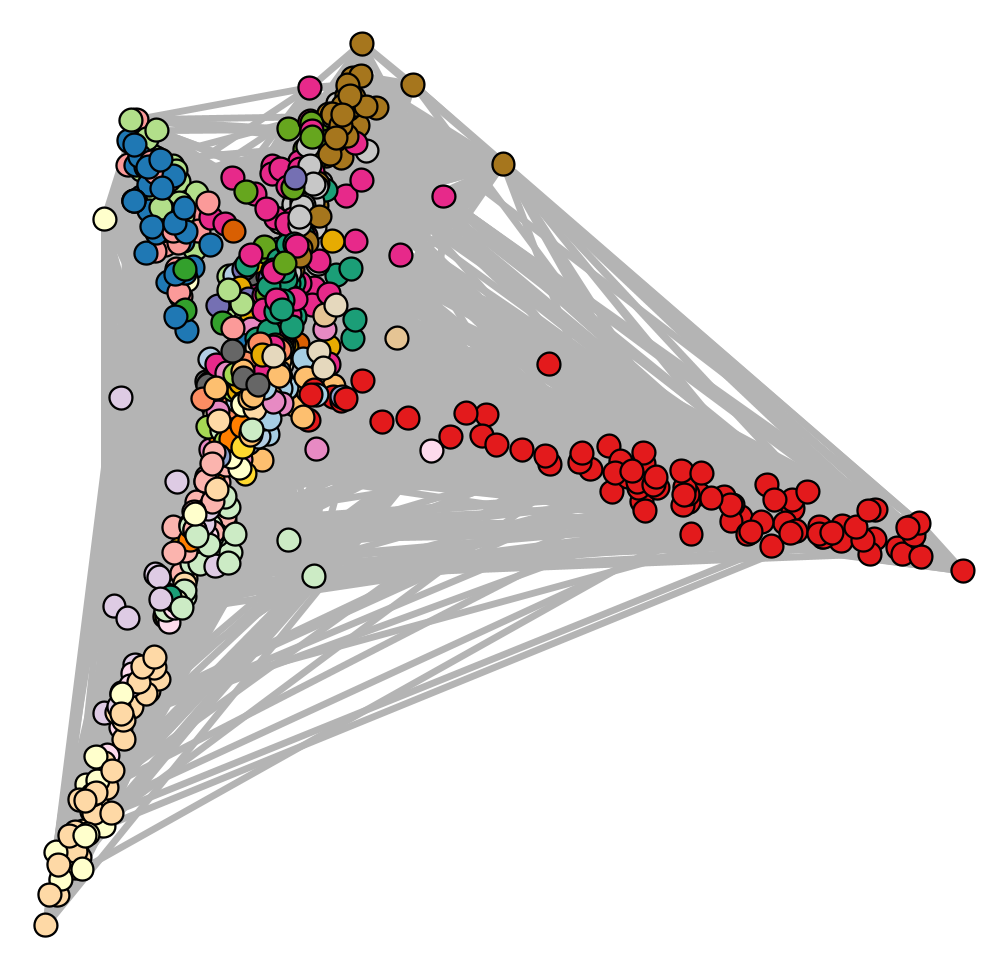} &
\includegraphics[width=0.22\textwidth]{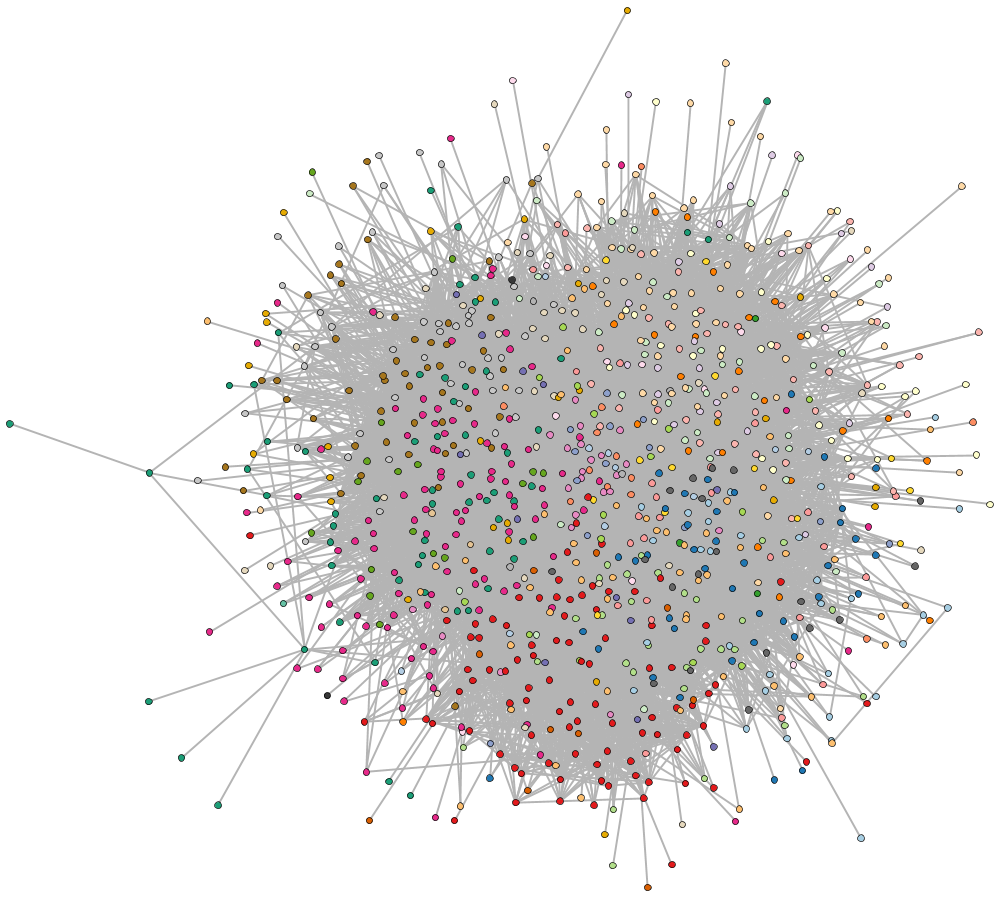} \\ \hline 
tsNET & Pivot MDS & sfdp & LinLog \\ \hline 
\includegraphics[width=0.22\textwidth]{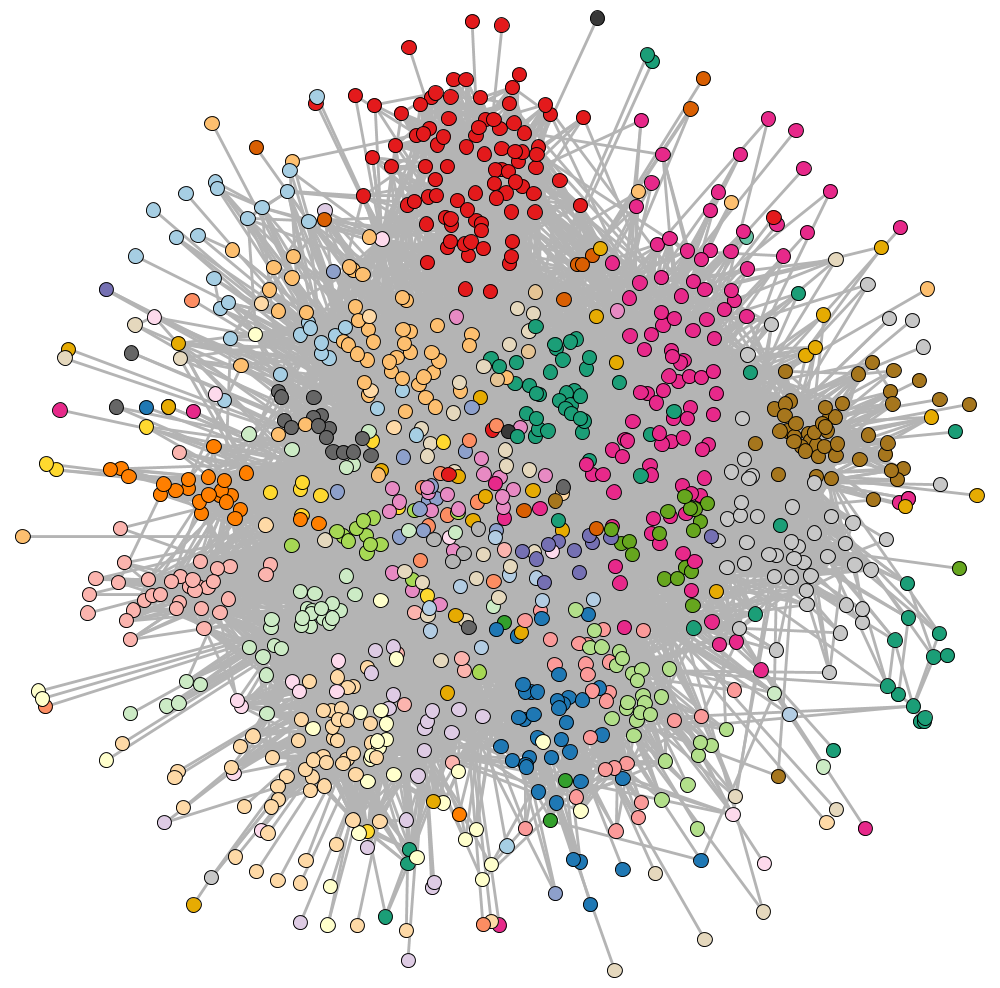} &
\includegraphics[width=0.22\textwidth]{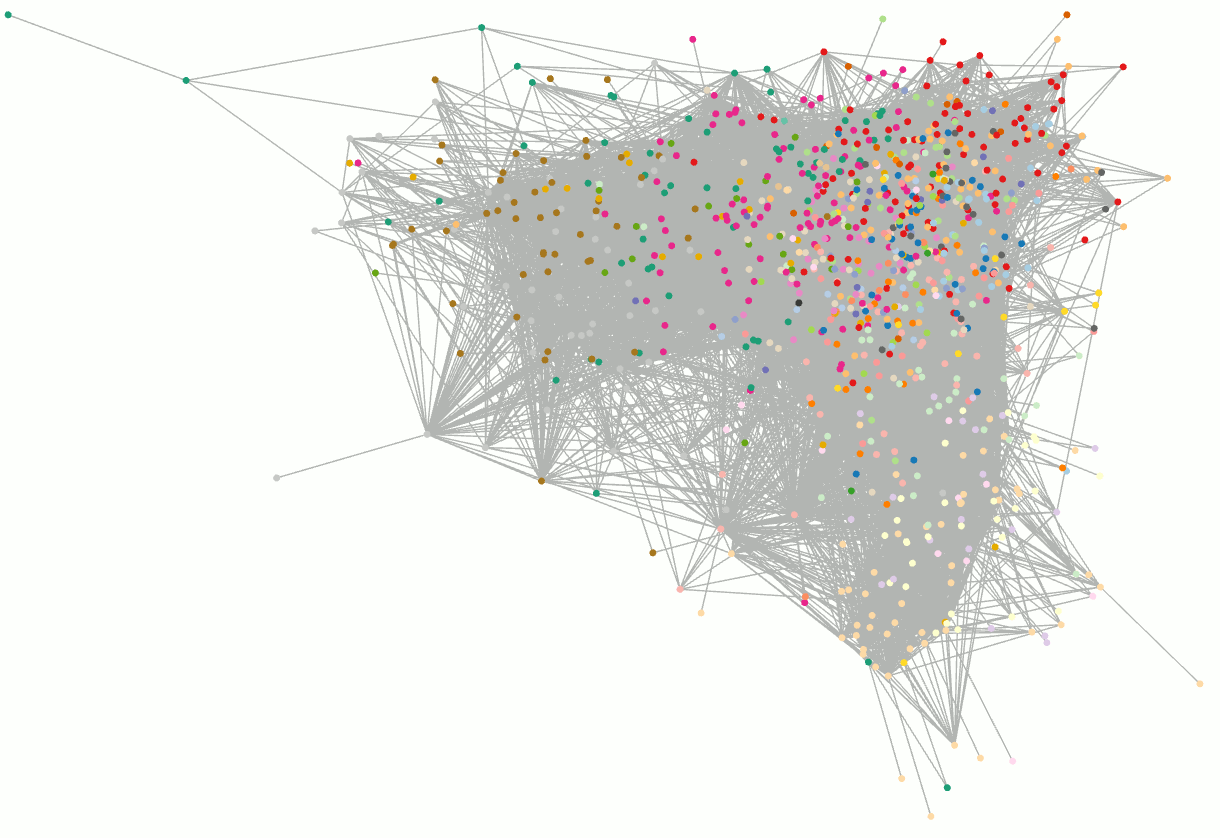} &
\includegraphics[width=0.22\textwidth]{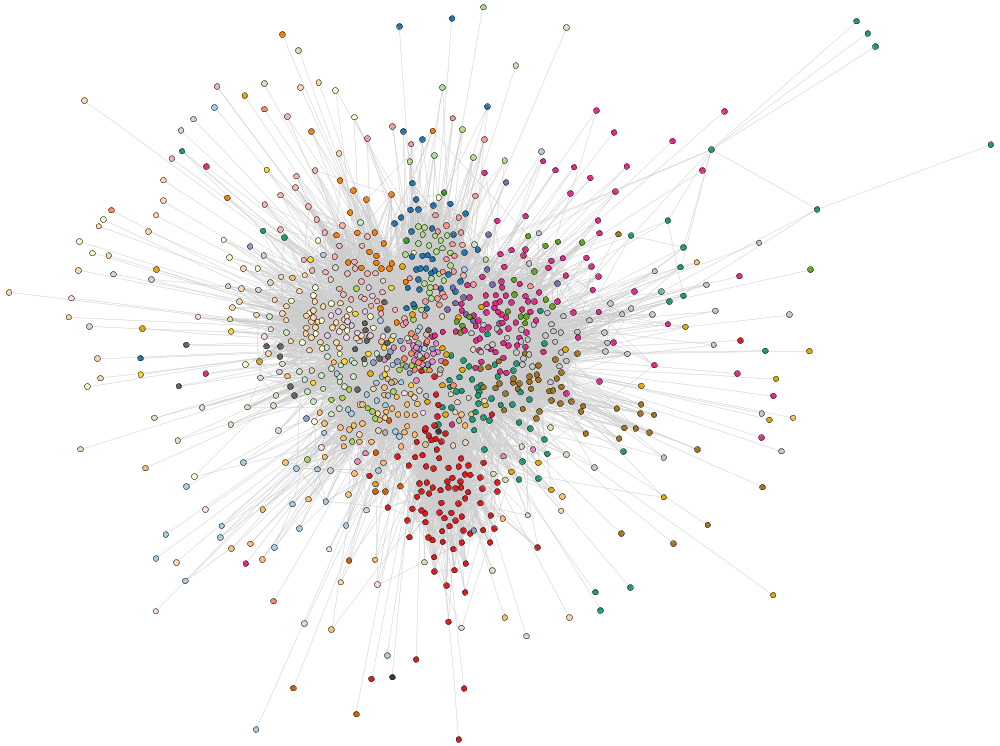} &
\includegraphics[height=0.22\textwidth]{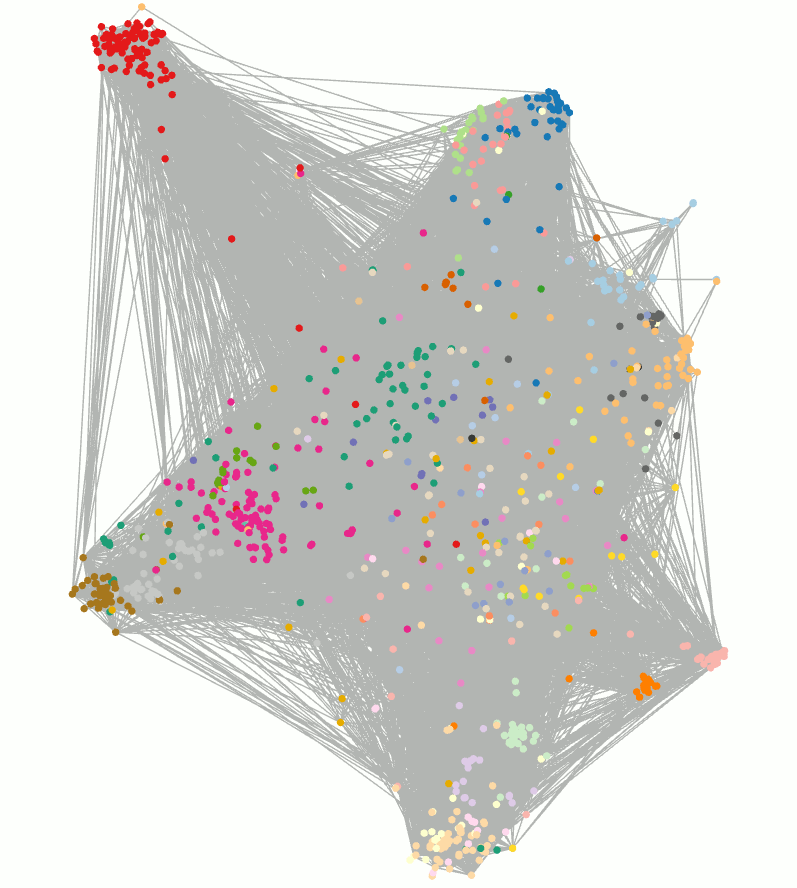} \\ \hline
\end{tabular}
\end{table}
\begin{figure}[H]
\centering
\includegraphics[width=0.8\textwidth]{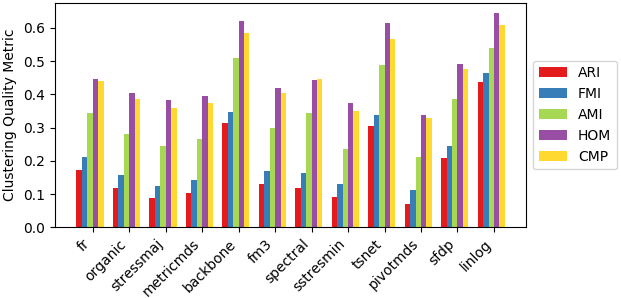}
\caption{Clustering quality metrics for \(email-Eu-core-lcc\). LinLog, backbone, and tsNET clearly outperform other layouts, as expected from Hypothesis 2. Among non-cluster-focused layouts, sfdp produces the highest scores.}
\label{fig:layoutcompmetrics_email}
\end{figure}

Tables \ref{table:layoutcomp_cfew} and \ref{table:layoutcomp_email} show layout comparison examples, with colours representing ground truth clusters, with \(CQ\) scores displayed in Figures \ref{fig:layoutcompmetrics_cfew} and \ref{fig:layoutcompmetrics_email} respectively. LinLog, tsNET, and Backbone score higher than other layouts for both datasets, supporting Hypothesis 2. In Table \ref{table:layoutcomp_cfew} and Figure \ref{fig:layoutcompmetrics_cfew}, where the number of clusters are small, other layouts such as sfdp, FR, FM3, and spectral also score close to 1. Meanwhile, in the example in Table \ref{table:layoutcomp_email} and Figure \ref{fig:layoutcompmetrics_email} displaying a real world graph with a larger number of clusters, LinLog, tsNET, and backbone's performances more clearly surpass the other layouts.
\begin{figure}[h]
\centering
\subfloat[Average for all datasets]{
\includegraphics[width=0.8\textwidth]{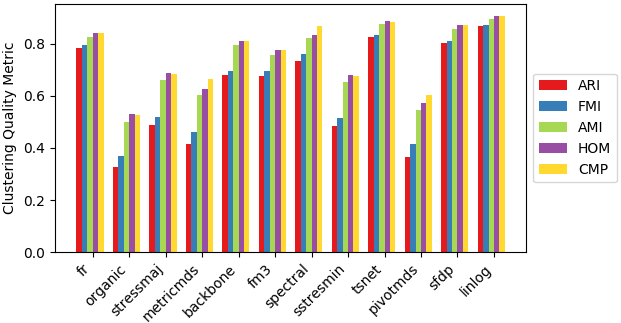}
}
\qquad
\subfloat[Average for real world datasets]{
\includegraphics[width=0.8\textwidth]{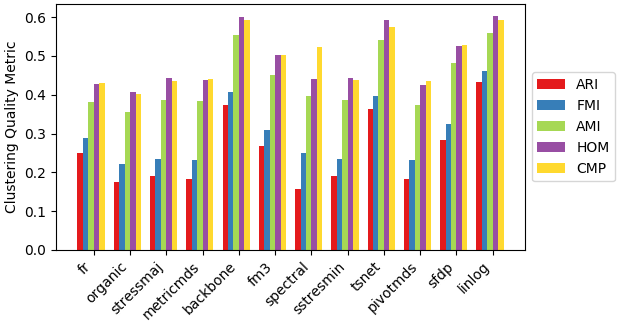}
}
\caption{Clustering quality metrics averaged per layout for all layout comparison datasets (a) and for real world datasets only (b). In (a), we see that tsNET and LinLog produce the highest scores, validating Hypothesis 2 for the two layouts. Meanwhile in (b), we see that on real world datasets, LinLog, tsNET, and Backbone outperforms other layout algorithms in accordance to Hypothesis 2.}
\label{fig:metrics_average}
\end{figure}

Figure \ref{fig:metrics_average} (a) shows the scores averaged across all layout comparison datasets and Figure \ref{fig:metrics_average} (b) show the scores averaged across real world datasets. Averaged across all datasets, LinLog scores the highest, with tsNET close behind, confirming Hypothesis 2 for these two layouts. Backbone scores well on many graphs, but sometimes deteriorates in quality when the number of clusters becomes larger compared to the total size of the graph, causing it to score lower than tsNET and LinLog on average (see Figure \ref{fig:metrics_average} (a)). Even so, it still outperforms the other algorithms on real world datasets as seen in Figure \ref{fig:metrics_average} (b), which supports Hypothesis 2 for Backbone on real world graphs.

In the case of synthetic datasets, sfdp also tends to perform well, as seen in the overall averaged clustering quality metric scores in Figure \ref{fig:metrics_average} (a). LinLog, backbone, and tsNET still outperforms it with real world datasets as seen from Figure \ref{fig:metrics_average} (b), however, in line with Hypothesis 2.

\subsection{Discussion and Summary}
\label{subsec:discussion}

Our experiments verify that LinLog and tsNET attains the highest average scores on our metrics across all comparison datasets and Backbone attains equally high average scores on real world datasets.

A point of note is that LinLog often has issues with excessive node overlaps, especially when the internal cluster density is high - this can be seen in Table \ref{table:layoutcomp_cfew}, where the nodes of each cluster are positioned very close together such that they almost appear as only one node, and to a lesser extent in Table \ref{table:layoutcomp_email} where the red cluster is packed quite closely together. Backbone does not have this problem on any tested graphs. Thus, we can conclude that Backbone also has its advantages for practical applications of clustered graph visualization.

\textit{In summary}, our experiments have confirmed Hypothesis 2 for LinLog and tsNET, which consistently obtained the highest scores across all datasets, while for Backbone it is more supported on real world structures.

\section{Conclusion and Future Work}
\label{sec:conclusion}

We have introduced a new graph drawing quality metric for the visualization of clusters in graph. Deformation experiments has shown the effectiveness of the metric in measuring how well a drawing of a graph depicts the clusters in the graph. We have also compared graph drawings produced by layouts emphasizing cluster structures to non-cluster-focused layouts and validated the claims of these cluster-focused layouts especially on real world structures.

A direction for future work is to refine the metric by combining it with readability metrics, such as to address node overlaps, and further validating it with human evaluation. Other geometric clustering algorithms besides \(k\)-means can also be tested, including fuzzy clustering algorithms that accomodate overlaps between clusters, and concepts of visual cluster separations for scatterplots~\cite{sedlmair2012taxonomy} can also be considered.

%
%
%
\bibliographystyle{splncs04}
%
\bibliography{clustering_GD}

\end{document}